\documentclass{llncs}
\usepackage{multirow}
\usepackage{amsmath}
\usepackage{amssymb}
\usepackage{floatflt}
\usepackage{graphicx}
\usepackage{booktabs} 
\usepackage[linesnumbered,algoruled,boxed,lined]{algorithm2e}
\SetKwInOut{Parameter}{parameter}

\DeclareMathOperator*{\argmax}{argmax}

\let\llncssubparagraph\subparagraph
\let\subparagraph\paragraph
\usepackage[compact]{titlesec}
\let\subparagraph\llncssubparagraph

\begin{document}
\title{Active Search for High Recall: a Non-Stationary Extension of Thompson Sampling }
\titlerunning{Active Search for High Recall} 
\author{Jean-Michel Renders}
\authorrunning{Jean-Michel Renders}
\tocauthor{Jean-Michel Renders}
\institute{Naver Labs Europe - 
\email{jean-michel.renders@naverlabs.com}}
\maketitle

\begin{abstract}
We consider the problem of Active Search, where a maximum of relevant objects –- ideally all relevant objects –- should be retrieved with the minimum effort or minimum time. Solving this kind of problem is crucial in applications such as fraud detection, e-discovery, prior art search in patent databases, \textit{etc}. Typically, there are two main challenges to face when tackling this problem: first, the class of relevant objects has often a very low prevalence and, secondly, this class can be multi-faceted or multi-modal: objects could be relevant for completely different reasons. To solve this problem and its associated issues, we propose an approach based on a non-stationary (aka restless) extension of Thompson Sampling, a well-known strategy for the Multi-Armed bandit problems. The collection is first soft-clustered into a finite set of components and a posterior distribution of getting a relevant object inside each cluster (or component) is updated after receiving the user feedback about the proposed instances. The``next instance'' selection strategy is a mixed, two-level decision process, where the algorithm first selects a cluster through ``optimistic Thompson sampling'' and then chooses ,inside the cluster, the instance with maximal relevance probability, as computed by an incremental on-line classifier. In some way, this method should be considered as an insurance, where the cost of the insurance is an extra exploration effort in the short run (i.e. the early stage of the search process), for achieving a nearly ``total'' recall with less efforts in the long run.
\end{abstract}

\section{Introduction}
Contrary to Active Learning, Active Search does not aim at building the best possible classifier with the minimum number of labelled instances, but simply aims at discovering virtually all the instances of the positive class (assuming a binary classification problem) with the minimum ``reviewing'' effort or cost. Active Search strategies most often include the incremental training of some classifier as a means to speed-up and to control the search, but this step is not strictly necessary. The collection -- or pool -- of objects to search in is assumed to be known in advance and the setting is, in some way, similar to the Transductive Learning setting, but with an on-line, incremental, ``recall-oriented'' perspective. Consequently, Active Search algorithms can be very different from traditional active learning algorithms. Active Search applications could be found in numerous domains: fraud detection, compliance monitoring, e-discovery, systematic medical reviews, prior art search when filing a patent, \textit{etc}.

Recently, some pieces of work \cite{Cormack2014,Cormack2016,DijkRKR15,ZhangLWCS15} have focused on developing Active Search strategies that significantly depart from active learning, by emphasizing the ``Total Recall'' aspect and the ``Continuous Active Learning'' setting.  The main idea of this family of works is to greedily select the next instances as the ones with the largest estimated probabilities of belonging to the relevant class (these probabilities are given by a classifier incrementally trained from all labelled instances up to the current time).   This selection strategy makes sense as these instances are both the  most promising ones -- given the data reviewed up to the current time -- and the ones that can reduce, at least locally and on the short run, the most the uncertainty over the parameters of the classifier. Interestingly enough, in \cite{GarnettKXSM12}, the authors give a bayesian rationale to this greedy strategy, showing that it maximizes the expected utility in the one-shot setting (i.e. when it remains only one possible query).

 In addition to belonging to a low prevalence class, relevant objects can take multiple forms or facets: the landscape of the positive class is often ``multi-modal''. When considering these challenges -- unbalanced class distribution and multi-modality of the relevant class --, there is a clear need to control the exploration-exploitation trade-off if we want to improve the ``baseline'' greedy approach and make it more robust: typically, the search starts with a small number of ``seed'' instances and these seeds rarely cover all modes and facets of the positive class. So, a greedy selection approach, which basically selects the next instances in the vicinity of the positive seeds through the classifier, runs the risk of missing large areas of relevant instances when the corresponding facets are not covered by (or hardly reachable from) the instances reviewed and labelled as positive up to the current time. At the early stage of the search, it can be useful to spend some effort in exploring diverse regions of the instance space, provided that these regions could offer potentially relevant elements in the long run. 

The use of Multi-Armed Bandits (MAB) appears as a natural choice to solve this exploration-exploitation trade-off. Instead of considering the problem as an instance recommendation problem with a binary response as in \cite{ChapelleMR14,ChapelleL11} and solving it by using contextual bandit strategies, we follow an alternative strategy that turns out to be more efficient in our use cases. This alternative consists in discretizing the structure of the feature space into a finite set of clusters and in relying on the cluster structure to manage the exploitation-exploration trade-off. More precisely, the idea is to consider each cluster as an arm of a MAB and to focus on the most promising ones, while ensuring that the selection strategy covers all facets or clusters of the instance space. However, we face the problem of dealing with non-stationary (or restless) mortal bandits (see for instance \cite{AllesiardoF15,garivier45,LevineCM17}). Indeed, each time the system selects a cluster (i.e. an arm), it consumes one instance inside the cluster and, as the cluster size is finite, the cluster can be exhausted. Moreover, even if the system chooses the most likely relevant instances inside a cluster, the reward distribution (the reward being 1 if the chosen instance is relevant and 0 otherwise) is clearly time-varying: roughly speaking, the chance to get extra relevant instances inside a cluster decreases over time. To deal with this non-stationary environment, the method proposed here introduces a forgetting factor when updating the reward posterior distributions of the arms (i.e. clusters).

\section{Related work}

Even if there is a large literature on active learning, the case of active search has not received the same attention. One of the first works on this problem was presented by \cite{GarnettKXSM12} who proposed a Bayesian approach that requires computations exponential in the number of lookahead steps (number of future steps considered), which makes this method impractical for large collections. Several pieces of work have approached active search as exploration on graphs \cite{WangGS13,MaHS15} or also graph learning \cite{Pfeiffer2014}. However, the mostly used state-of-the-art method remains the greedy (one single step ahead) approach: one of the most striking examples of this is the ``Continuous Active Learning'' concept of \cite{Cormack2016}, implemented in the form of the AutoTAR system in the field of Technology-assisted Review for e-Discovery \cite{Cormack2016} (extended further in \cite{ZhangLWCS15}). Our method is quite different from these active  search approaches, as it explicitly controls the exploration-exploitation trade-off, does not use any graph structure, and does not rely on a finite-horizon approach.

Our proposed method shares some similarity with introducing diversity in recommendation, retrieval and active learning: instead of proposing the instance predicted as the most relevant (or the most uncertain) given the feedback received up to the current round, the selection criterion involves an extra term based on diversity (see for instance \cite{XuAZ07} and, more fundamentally, the concept of Maximal Marginal Relevance ranking \cite{Carbonell98}). In our method, we are promoting diversity in a different way, by exploiting the cluster structure of the data, and by using MAB in a suitable way.

Our approach to solve active search builds on ideas of cluster-based active learning, based on the ``cluster assumption'' (instances indside the same cluster are likely to have the same label. \cite{Nguyen2004} and \cite{Hong2015} approached active learning by clustering unlabeled instances and selecting instances to query according to a criterion that prefers clusters on the decision boundary and the most representative instances in these clusters.

The use of MAB in Active Learning is not new and relatively well studied. A realtively common way to solve active learning with MAB is to cluster the instances on the pool and consider that each cluster is an arm \cite{Bouneffouf2014,Collet2015}. In this case, the payoff distribution for each arm is non-stationary since the probability of finding relevant instances in a cluster decreases as the cluster is exploited. The approach of \cite{Bouneffouf2016}, for instance, deals with the non-stationarity by assuming a known fixed trend. 

Our work differs from these veins of works by different factors: first of all, these methods tackle the Active Learning problem, while we are targeting the Active Search problem. Secondly, they often rely on a simple classification model, namely that instances inside a cluster all have the same posterior probability of being relevant. Thirdly, we use MAB to explicitly control the exploration-exploitation trade-off. Lastly, we use soft clustering, that allows to propagate acquired information outside a cluster and to speed-up the search.

Finally, our work is not the only one to use MAB for Active Search. Even if initially formulated as an item recommendation problem, the MAB-based approach of \cite{ChapelleL11,ChapelleMR14} could be used as such for active search; in this work, the ``next item selection'' problem is expressed as a contextual bandit, where each instance is an arm; the expected reward (or relevance label) of an instance is expressed as a logistic regression model, whose parameters are sampled following a Thompson sampling strategy from the posterior distribution updated each time a new label is collected. We have implemented this method for our collections (see experimental section) and the results were extremely weak due to the fact that the method is not adapted to sparse high-dimensional data and requires a lot of exploration, often exceeding the budget or simply the dataset size.

\section{Proposed Method} 

Traditional active search methods are typically greedy: they are looking for 
the most promising instances, i.e. the ones with the highest conditional 
probabilities ($\pi_i = p(Y_i=1|x_i)$) as estimated by a classifier trained on 
the instances reviewed up to the current time. As the class of relevant 
instances often has a very low prevalence and could be considered as a 
``rare event'', strategies of this kind make sense as these instances are actually the ones 
minimizing the risk of reviewing a non-relevant instance while, simultaneously, 
reducing the uncertainty as much as possible. For instance, with a logistic regression 
classifier, most of the instances in the poolset will have probability scores 
$\pi_i$ not larger than the prevalence and only a few will have a probability 
larger and closer to 0.5 (\cite{kingzeng01} provides details 
about these observations). Entropy-based or uncertainty-based active learning 
strategies will favour these instances. Note that this also corresponds to the 
instances that are the most statistically informative with respect to the 
variance matrix on the parameter estimates, as an instance contributes to the 
inverse of this matrix through a factor $\pi_i (1-\pi_i)$  \cite{kingzeng01}.

However, greedy strategies are likely to fail when the relevant class is 
``multi-modal'' or ``multi-faceted'', in other words when they are multiple, well 
distinct and possible unbalanced ways of being relevant. As the greedy 
strategies introduce a high selection bias when incrementally building the 
training set, it could be that the selection algorithm will miss important 
sectors of the relevant category, because they are relative far from the 
positively-labelled training instances, which constitute an homogeneous set by 
construction. In practice, this risk strongly depends on the quality of the 
seed set: the seed set should be diverse enough and have a good coverage of 
the different facets of relevance, but unfortunately this 
guarantee is hard to obtain. 

There are several ways of translating the active search problem into a MAB 
problem. For instance, each instance could be considered an an arm but once it 
is selected, it doesn't make any sense to choose this arm again. The alternative we 
adopt in this work, is to choose clusters of pool instances as arms. But this 
choice has the particularity that the algorithm can ``exhaust'' a cluster 
and that, consequently, the related arm will ``die'' once every instance in the 
cluster have been reviewed and labelled. If we consider the reward of an arm as 
the binary label of the instance that the algorithm will choose inside the 
corresponding cluster (1 if the instance is relevant and 0 otherwise), the 
reward distribution of an arm is obviously non-stationary as an ``exhausted'' 
empty cluster will switch abruptly towards a single-value discrete distribution 
(0 reward with probability 1). Intuitively, we face a diminishing return issue: the retrieval rate of relevant objects decreases as we are exploiting the cluster.  Note that, in our approach, we rely on soft clustering
so that an instance can belong to multiple clusters with different degrees of membership. This renders the approach more robust with respect to a particular clustering method but, on the other side,  we have to adapt the MAB algorithms for a non-standard reward scheme: the reward obtained for a particular instance should be re-assigned to multiple clusters (or arms) with an appropriate weighting.

It is rather usual, in active search, to proceed by batches of (polynomially or 
exponentially)increasing size: instead of proposing one single instance to 
review at each iteration, the active search algorithm provides batches of 
instances to be reviewed which are larger and larger, as the confidence about 
the classifier performance becomes higher \cite{Cormack2016}. 
We also have to adapt the MAB algorithms so that they can provide us with a 
``batch'' of arm trials -- some kind of MAB with multiple plays -- and update 
the arms' sufficient statistics (i.e. the posterior distribution of the reward) accordingly.

Basically, our algorithm consists of the following steps:

(1) Create an initial training set from the seed set, consisting
of a synthetic ``positive'' instance built from the initial user query or of a few positive 
instances discovered by any means, and a random sample from the poolset, 
temporarily labelled as ``negative'' (typically, the random sample size is 100 
instances). As the relevant class has low prevalence, most of the instances in 
the random sample are indeed non-relevant and the label noise introduced by this 
approximation is negligible.

(2) Iteratively, create a batch of instances from the pool set using a 
non-stationary, batch (or multiple-plays) extension of the ``Thompson sampling'' MAB 
algorithm, ask the reviewer to label them, remove them form the pool set, update 
the reward distribution estimates; and retrain the 
classifier based on the labelled documents and a random sample re-drawn from the 
pool set, temporarily labelled as negative.

Actually, the MAB algorithm is a two-level process, where the algorithm first samples $B$ times ($B$ being the batch size) a ``conversion rate'' for each cluster/arm (the conversion rate is the probability of a high-score member of the cluster to be annotated as a relevant instance) from a Beta posterior distribution and, secondly, selects an instance that maximizes the probability of being relevant, given the sampled conversion rates of the clusters it belongs to.

The Thompson sampling extension, to deal with non-stationary reward of the 
``mortal'' multi-armed bandits with multiple plays (batch of trials), is based on the 
following ideas:
\begin{enumerate}
\item The posterior distribution of the ``conversion rate'' of each arm/cluster is a Beta distribution with 
parameters ($S_k,F_k$), initialised with $S_k=F_k=0.5$ for all arms $k$ 
(Jeffreys's prior). The parameters $S_k$ and $F_k$ could be considered as the 
``equivalent'' number of successes and failures of a binomial distribution (the  
cluster reward distribution).
\item When receiving the label (or, equivalently, the reward) of the instances selected at the previous round, the binary reward is re-distributed over the the clusters with a weight equal to the membership value of the instance with respect to the cluster. 
\item  Updating the posterior is done using a forgetting factor, discounting the 
previous (weighted) success/failure counts by a factor $\gamma$. Alternatively, it can be 
done using a sliding window of size $W$; in this case, only the (weighted) success/failure 
counts of the last $W$ iterations are taken into account in updating the 
posterior distributions.
\item For a batch of size $B$, we repeat $B$ times the following steps: for each 
arm/cluster, draw a value $\theta_{k}$ from the Beta distribution associated to 
the cluster: $\theta_k \propto Beta(S_k,F_k)$; this value should be interpreted as 
the parameter (the mean) of a Bernoulli distribution modelling the arm reward 
distribution; in this work, we use an ``Optimistic Bayesian sampling'' variant, where one does not allow 
$\theta_k$ to be smaller than the empirical discounted mean of the arm reward (based on observations up to the current round).
\item When generating the batch of instances, once an instance has been selected, it is removed from the pool set and will not be selected again.
\end{enumerate}

Let us be more precise in the description of the update of the cluster posterior distribution. Assuming that, at round $t$, the conversion rate of cluster $k$ (i.e. the probability that a ``high score'' member of the cluster $k$ will be annotated as relevant) follows a Beta distribution with parameters ($S^{(t)}_k,F^{(t)}_k$), then the posterior distribution $\theta_k $ of the conversion rate at round $(t+1)$ is $Beta(S^{(t+1)}_k,F^{(t+1)}_k)$ with:
\begin{eqnarray}
  S^{(t+1)}_k &=& \gamma S^{(t)}_k + \sum_{i \in B^{(t)}} r_i \mu_{i,k} \nonumber \\
  F^{(t+1)}_k &=& \gamma F^{(t)}_k + \sum_{i \in B^{(t)}} (1-r_i) \mu_{i,k} \label{eq:update_equations}
\end{eqnarray}
In these equations, $\gamma$ is the forgetting factor to cope with the non-stationarity of the conversion rate distribution, $B^{(t)}$ is the batch of instances selected at round $t$, $\mu_{i,k}$ is the membership value of $i$ in cluster $k$ (interpreted as $p(k|x_i)$, the probability that instance $i$ with feature vector $x_i$ belongs to cluster $k$) and $r_i$ is the binary reward (i.e. 0/1 label) of instance $i$.

As far as the selection criterion is concerned, at each round $t$, we repeat $B$ times the following steps: for each cluster $k$, sample $\theta_{k}$ from $Beta(S^{t}_k,F^{t}_k)$ and compute $\theta^*_{k} = \max (\frac{S^{t}_k}{S^{t}_k+F^{t}_k},\theta_{k})$ (optimistic Thompson sampling, replacing the sampled value by the empirical mean if the former is smaller than the latter); then choose the instance $i^*$ such that:
\begin{equation}
\label{eq:selection_criterion}
 i^* = \argmax_{i \in U} \pi_i^{(t)} \sum_k \mu_{i,k} \theta^*_k
\end{equation}
with $U$ the set of unlabelled instances (i.e. the pool), $\pi_i^{(t)}$, the probability that instance $i$ is relevant, as estimated by the current classifier using labelled instances up to round $t$. Intuitively, this criterion selects the instance that has the best ``optimistic'' chance of being converted towards a real relevant instance: this ``chance'' is measured by the product of the marginal ``optimistic'' conversion rate of ``high score'' instances (marginalised over the clusters the instance belongs to) and, roughly speaking, the probability of being a ``high score'' instance (which is trivially estimated by $\pi_i^{(t)}$). The exploration effect relies on the sampling from $Beta(S^{t}_k,F^{t}_k)$, which can potentially favour less explored clusters as their posterior distribution is less peaked.

Concretely, the method we propose is summarised by Algorithm \ref{alg:main_algo}. 

\begin{algorithm}
\label{alg:main_algo}
\SetKwData{Left}{left}
\SetKwData{This}{this}
\SetKwData{Up}{up}
\SetKwFunction{Union}{Union}
\SetKwFunction{FindCompress}{FindCompress}
\SetKwInOut{Input}{input}
\SetKwInOut{Output}{output}
\Input{Collection of $n$ instances $D=L \cup U =\{x_1, ...,x_n\}$, grouped into $K$ soft clusters
$C_{1,...,K}$ with memberships $\mu_{i,k}=p(C_k|x_i)$ ($i=1,..,n$,$k=1,...,K$), a seed set of labeled examples $L=\{ (x_1,y_1),...,(x_s,y_s)  \}$, 
a pool set of unlabeled documents $U=\{x_{s+1},...,x_n  \}$, Budget $G$, 
pseudo-negative random sample size $n$, forgetting factor $\gamma$}
\Output{List of relevant instances}
\BlankLine
- Create a training set $T$ from the union of $L$ and a random sample of $n$ 
instances from $U$ (temporarily) labelled as negative\;
- Train a classifier $C$ from $T$ and associate a relevance score $\pi^0_i$ to 
each document $i$ of $U$\;
- $B \leftarrow 1$\;
- $t \leftarrow 0$\;
- Initialise the parameters of the Beta distributions associated to each 
arm/cluster with Jeffreys' priors ($S^0_k=F^0_k=0.5$ for $k=1,\ldots\,K$)\;
\While{$B\leq G$}{
- Initialise Batch $W \leftarrow \emptyset$\;
 \For {$b \leftarrow 1$ \KwTo $B$}{
 - Sample $\theta_{k}$ from $Beta(S^t_k,F^t_k)$ ($k=1,\ldots,K$)\;
 - Compute $\theta^*_{k} = \max(\theta_{k}, \frac{S^{t}_k}{S^{t}_k+F^{t}_k})$\;
 - Select the instance $i^*$ that maximizes equation \ref{eq:selection_criterion}\;
 - Push $i^*$ in the batch: $W \leftarrow L \cup \{i^*\}$\;
 - Remove $i^*$ from the pool of unlabelled instances $U$\;
 }
- Ask the reviewers to label the batch $W$, add the labelled instances to $L$ and segregate the relevant instances in the output list 
\;
- $B \leftarrow B + \lceil \frac{B}{40} \rceil$\;
- $t \leftarrow t + 1$\;
- Update the $S_k$ and $F_k$ parameters of the Beta distributions associated to 
each arm/cluster by using the labels (rewards) of the documents in $W$ and the 
forgetting factor $\gamma$, as described in equations \ref{eq:update_equations}\;
- Create a new training set $T$ from the union of $L$ and a random sample of $n$ 
documents from $U$ (temporarily) labelled as negative\;
- Retrain a classifier $C$ from $T$ and associate a relevance score $\pi^t_i$ to 
each document $i$ of $U$\;
}
\caption{Active Search with MAB}\label{algo_disjdecomp}
\end{algorithm}

\section{Experiments}

\subsection{Datasets}
We have chosen two representative datasets that contain multi-modal low-prevalence classes. Both are collection of text documents, but the method is not restricted to textual data (we are currently applying it to movie recommendation problems, based on previous ratings and on item/user features). The first dataset is the Reuters RCV1 Corpus (approximately 807,000 documents), considering only classes at the first level of the hierarchy with a prevalence less than 10\% and having at least three children sub-classes whose cumulated size is at least 50\% of the parent class. These classes are: \textbf{C17} (Funding/Capital - 5.18\% prevalence), \textbf{C18} (Ownership Changes - 6.38\% prevalence), \textbf{E14} (Consumer Finance - 0.26\% prevalence), \textbf{E31} (Output/Capacity - 0.29\% prevalence), \textbf{E51} (Trade/Reserves - 2.57\% prevalence), \textbf{G15} (European Community - 2.37\% prevalence). The reason of selecting these classes is that, by construction, they consist of multiple diverse sub-classes and, consequently, are ``multi-faceted''.

The second collection is the ENRON collection and a set of 8 associated ``topics'' (i.e. a set of 8 different relevance classes). These 8 topics are the ones used in the TREC 2010 Legal Track benchmark (Topics 200 to 207: see \cite{CormackGHO10} for details on these topics). These topics have prevalence well behind 0.2\% (0.1\% on average).  We conjecture that, as in many realistic e-discovery tasks, these topics are also multi-modal/multi-faceted. As most of the corpus is not labelled with respect to these 8 classes, we considered the subset (called subsequently ``ENRON-subset'') of document assessed for at least one of these topics (resulting in approximately 60,000 documents). In  the TREC Legal Track benchmarks, the selection of the documents to be assessed for each topic is done by stratified sampling, the strata being defined by the degree of ``agreement'' on the relevance by the different systems/teams participating to the benchmark. In particular, the last stratum typically consists of documents never returned by any participating system and, even if it constitutes the main part of the collection, this stratum has a low sampling rate. All performance measures take into account this stratified sampling process by weighting each document with the inverse of the sampling rate of the stratum it belongs to. This means that retrieving a relevant document in the low stratum part is considered as more important (and, consequently rewarded much more) than retrieving relevant documents in the first stratum -- the stratum of ``easy to find'' relevant documents, as any participating system discovered them. In the results shown here after, we have taken these stratum weights into account in computing the achieved recall.

\subsection{Results}

The classifier used in all experiments is a L2-regularised Logistic Regression, based on the tokenised bag-of-word representation of the collections (stop-words included, but words with document frequency less than 3 filtered out); TF-IDF weighting scheme and L2-normalisation were applied to each document vector. As soft clustering method, we used LDA (Latent Dirichlet Analysis). 

In both cases, we fixed the discounting factor $\gamma$ (in updating the priors of the cluster reward distribution) to 0.99 for Reuters RCV1, and to 0.95 for ENRON-subset. The number of arms/clusters ($K$) -- or, equivalently, the number of latent components in LDA -- was fixed to 200 for Reuters RCV1 and to 50 for ENRON-subset. These hyper-parameters were tuned on ``unused'' topics: the ``Commodity Markets'' class (\textbf{M14}) of Reuters RCV1 and the topic 301 of the TREC 2010 Legal Interactive Task.

The performance measure is simply the proportion of the collection to be reviewed to reach certain levels of recall, focusing on the high recall values. For each collection and each topic, we have performed 10 different runs with different seed sets; each seed set consisted of three random relevant instances of the class or topic. We have limited the reviewing budget  to 40\% of the collection. The values given in the tables \ref{tab:ENRON_res} (for the ENRON-subset dataset) and \ref{tab:RCV1_res} (for the Reuters RCV1 dataset) are the average over these 10 runs.

\begin{table}[htbp]

  \centering
  \caption{Collection 1: ENRON-subset. Percentage of the collection to be reviewed to reach a Recall level. As the budget is limited, ``$>$40\%'' means that the budget was exhausted before reaching the desired recall level. }
  \resizebox{\textwidth}{!}{
    \begin{tabular}{|l|rrrr|rrrr|}
     \hline
    TOPIC & \multicolumn{4}{c|}{Baseline} & \multicolumn{4}{c|}{Proposed Method} \\
          & \multicolumn{1}{l}{Recall=0.5} & \multicolumn{1}{l}{Recall=0.85} & \multicolumn{1}{l}{Recall=0.90} & \multicolumn{1}{l|}{Recall=0.975} & \multicolumn{1}{l}{Recall=0.5} & \multicolumn{1}{l}{Recall=0.85} & \multicolumn{1}{l}{Recall=0.90} & \multicolumn{1}{l|}{Recall=0.975} \\\hline\hline
    200   & 0.29\%  & 2.34\%  & 2.51\%  & 4.86\%  & 0.31\%  & 2.41\%  & 2.52\%  & 3.88\% \\\hline
    201   & 1.01\%  & 2.23\%  & 2.36\%  & 3.16\%  & 0.74\%  & 1.87\%  & 1.99\%  & 2.97\% \\\hline
    202   & 0.55\%  & 4.9\%   & 6.68\%  & 10.3\%  & 0.54\%  & 4.48\%  & 6.23\%  & 8.23\% \\\hline
    203   & 0.86\%  & $>$40\%   & $>$40\%   & $>$40\%   & 0.99\%  & 13.52\% & 13.61\% & 13.8\% \\\hline
    204   & 1.45\%  & 8.26\%  & 10.45\% & 14.81\% & 1.41\%  & 7.64\%  & 9.88\%  & 13.36\% \\\hline
    205   & 4.38\%  & 18.2\%  & $>$40\%   & $>$40\%   & 4.78\%  & 18.8\%  & 19.4\%  & $>$40\% \\\hline
    206   & 1.07\%  & $>$40\%   & $>$40\%   & $>$40\%   & 1.15\%  & 11.6\%  & 14.26\% & $>$40\% \\\hline
    207   & 0.29\%  & 1.68\%  & 2.42\%  & $>$40\%   & 0.32\%  & 1.71\%  & 2.2\%   & 3.89\% \\
     \hline
    \end{tabular}}
  \label{tab:ENRON_res}%
\end{table}%

\begin{table}[htbp]
  \centering
  \caption{Collection 2: Reuters RCV1. Percentage of the collection to be reviewed to reach a Recall level}
    \begin{tabular}{|l|rrr|rrr|}
    \hline
    TOPIC & \multicolumn{3}{c|}{Baseline} & \multicolumn{3}{c|}{Proposed Method} \\
          & \multicolumn{1}{l}{Recall=0.9} & \multicolumn{1}{l}{Recall=0.95} & \multicolumn{1}{l|}{Recall=0.99} & \multicolumn{1}{l}{Recall=0.9} & \multicolumn{1}{l}{Recall=0.95} & \multicolumn{1}{l|}{Recall=0.99} \\\hline\hline
    C17   & 7.92\%  & 12.27\%  & 25.97\%  & 8.15\%   & 11.9\%   & 19.48\%  \\\hline
    C18   & 6.81\%   & 8.96\%   & 15.87\%  & 6.91\%   & 8.75\%   & 11.21\%  \\\hline
    E14   & 1.19\%   & 2.99\%   & 12.63\%  & 1.48\%   & 2.81\%   & 10.71\%  \\\hline
    E31   & 0.93\%   & 1.67\%   & 17.06\%  & 1.13\%   & 1.61\%   & 12.34\%  \\\hline
    E51   & 6.67\%   & 10.42\%  & 19.85\%  & 6.81\%   & 10.01\%  & 13.45\%  \\\hline
    G15   & 2.39\%   & 2.98\%   & 5.36\%   & 2.5\%    & 2.92\%   & 4.12\%  \\
    \hline
    \end{tabular}%
  \label{tab:RCV1_res}%
\end{table}%

There are several important observations that we can make from these experimental results:
\begin{enumerate}
\item If the requested level of recall is relatively low, the baseline is still the best choice. But, for a sufficiently high recall, the exploration effort spent during the early phases of the search starts to be beneficial and our method outperforms the baseline. The ``break-even'' point between the two strategies depends on the collection and on the topic. In some way, our method can be considered as an ``insurance'' to be able to reach efficiently a high recall without forgetting significant segments of relevant instances; the cost of this insurance is the extra effort spent in exploring diverse clusters during the search.
\item The beneficial effect seems to decrease, and even to disappear, for extreme values of recall, especially in the case of the ENRON corpus. The most likely reason of this sudden decline is the label noise: some irrelevant instances -- incorrectly labelled as relevant -- are virtually unreachable from any classifier built from (correctly labelled) positive instances. And we suspect that noisy labels are more numerous with the ENRON dataset (the relevant class is more ``fuzzy'' and subject to interpretation) than with the Reuters corpus.
\end{enumerate}

\section{Conclusions and Future Works}
This paper considers the Active Search problem as a resource allocation task in an uncertain environment and handles it in a way similar to what is done for petroleum drilling and ore mining projects. By soft-clustering the landscape of the instance feature space and using sampling strategies based on MAB, the method proposed here should be considered as an insurance to decrease the risk of missing a significant amount of relevant objects when the task is to achieve high recall of a low-prevalence, multi-faceted relevant class. Future works will focus on analysing the cost/benefit ratio depending on the task and the collection to be processed. 

\bigskip

\textbf{Aknowledgment}: This work was partially funded by the French Government under the grant   $<$ANR-13-CORD-0020$>$ (ALICIA Project).

\bibliographystyle{plain}
\bibliography{ECIR2018}

\end{document}